\documentclass[12pt]{article}

\usepackage{newtxtext,newtxmath}

\usepackage{bm}
\usepackage{xcolor}
\usepackage{graphicx}
\usepackage{booktabs}
\usepackage{multirow}
\usepackage{makecell}
\usepackage{array}
\usepackage{siunitx}
\usepackage{threeparttable}
\usepackage{url}
\usepackage{scicite}
\usepackage[hidelinks]{hyperref}

\usepackage[letterpaper,margin=1in]{geometry}

\renewenvironment{abstract}
	{\quotation}
	{\endquotation}

\date{}

\makeatletter
\renewcommand{\fnum@figure}{\textbf{Figure \thefigure}}
\renewcommand{\fnum@table}{\textbf{Table \thetable}}
\makeatother

\def\scititle{
	An All-van-der-Waals Qubit
}
\title{\bfseries \boldmath \scititle}
\usepackage{authblk}

\author{
    Sein~Park$^{1}$,
    Sameia Zaman$^{1,2}$,
    Junghyun Kim$^{1,2}$,
    Junyoung An$^{1,2}$,
    \and
    Daniel~Rodan-Legrain$^{1}$,
    Hung-Yu~Tsao$^{1,2}$,
    Chia-Chin~Tsai$^{1,3}$,
    Aranya~Goswami$^{1}$,
    \and
    R\'eouven~Assouly$^{1}$,
    William~P.~Banner$^{1,2}$,
    Gabriel~D.~Cutter$^{1,2}$,
    \and
    Kenji~Watanabe$^{4}$,
    Takashi~Taniguchi$^{5}$,
    Terry~P.~Orlando$^{1,2}$,
    Gil-Ho~Lee$^{6}$,
    \and
    Kyle~Serniak$^{1,7}$,
    Max~Hays$^{1}$,
    Jeffrey~A.~Grover$^{1}$,
    Philip~Kim$^{8,9}$,
    \and
    Pablo~Jarillo-Herrero$^{1,10}$,
    Joel~\^I-j.~Wang$^{1,11,\ast}$,
    
    William~D.~Oliver$^{1,2,10,\ast}$
    \and
    \affil[1]{Research Laboratory of Electronics, Massachusetts Institute of Technology, Cambridge, MA 02139, USA}

	\small$^{1}$Research Laboratory of Electronics, Massachusetts Institute of Technology, Cambridge, MA 02139, USA\and
	\small$^{2}$ Department of Electrical Engineering and Computer Science, Massachusetts Institute of Technology,\and \small Cambridge, MA 02139, USA\and
    \small$^{3}$Department of Materials Science and Engineering, Massachusetts Institute of Technology,\and \small Cambridge, MA 02139, USA\and
    \small$^{4}$Research Center for Electronic and Optical Materials, National Institute for Materials Science, 1-1 Namiki,\and \small Tsukuba 305-0044, Japan\and
    \small$^{5}$Research Center for Materials Nanoarchitectonics, National Institute for Materials Science, 1-1 Namiki,\and \small Tsukuba 305-0044, Japan\and
    \small$^{6}$Department of Physics, Pohang University of Science and Technology, Pohang, Gyeongsangbuk-do 37673,\and \small Republic of Korea\and
    \small$^{7}$Lincoln Laboratory, Massachusetts Institute of Technology, Lexington, MA 02421, USA\and
    \small$^{8}$John A. Paulson School of Engineering and Applied Sciences, Harvard University, Cambridge, MA, USA\and
    \small$^{9}$Department of Physics, Harvard University, Cambridge, MA, USA\and
    \small$^{10}$Department of Physics, Massachusetts Institute of Technology, Cambridge, MA 02139, USA\and
    \small$^{11}$Department of Physics, New York University, New York, NY 10003, USA.\and
	\small$^\ast$Corresponding author. Email: joelwang@nyu.edu and william.oliver@mit.edu\and
}

\begin{document} 


\maketitle

\begin{abstract} \bfseries \boldmath
Advances in solid-state physics, materials science, and device engineering have accelerated the development of superconducting qubits~\cite{Nakamura1999CoherentControl,Koch2007Transmon,Schuster2007PhotonNumber,Houck2008Transmon,Schreier2008Suppressing,Xmon2013,clerk2020hybrid,Bland2025Millisecond}. Among emerging platforms, van der Waals (vdW) materials and their heterostructures are potentially attractive building blocks for quantum devices, yet their realization in qubit architectures remains underexplored~\cite{wang2019coherent,wang2022hexagonal,Antony2021hBN,Jesse2026WSe2,Balgley2025Coherent,DElia2025Coherent}. Here we report an all-vdW superconducting qubit based on a NbSe$_2$--hBN--NbSe$_2$ junction, in which a thin hBN layer simultaneously provides Josephson coupling and capacitive shunting between two NbSe$_2$ leads, forming a ``merged-element" transmon. Temporal characterization using circuit quantum electrodynamics (cQED) techniques yields an average energy-relaxation time $T_{1,\mathrm{avg}} = 55 ~\si{\micro\second}$, Hahn-echo coherence time $T_{2\mathrm{E},\mathrm{avg}} = 21~\si{\micro\second}$, and Ramsey coherence time $T_{2\mathrm{R},\mathrm{avg}} = 1.9~\si{\micro\second}$. The relatively low Ramsey time is primarily attributable to an enhanced sensitivity to charge noise consistent with the realized device parameters and not a fundamental limitation. These results show that lumped-element superconducting qubits based on vdW heterostructures can achieve coherence times comparable to those of conventional Al--AlO$_\mathrm{x}$--Al qubits, while offering a reduced device footprint and suppressed stray capacitive coupling. 


\end{abstract}
\begin{figure}[htbp]
    \centering
    \includegraphics[width=0.8\linewidth]{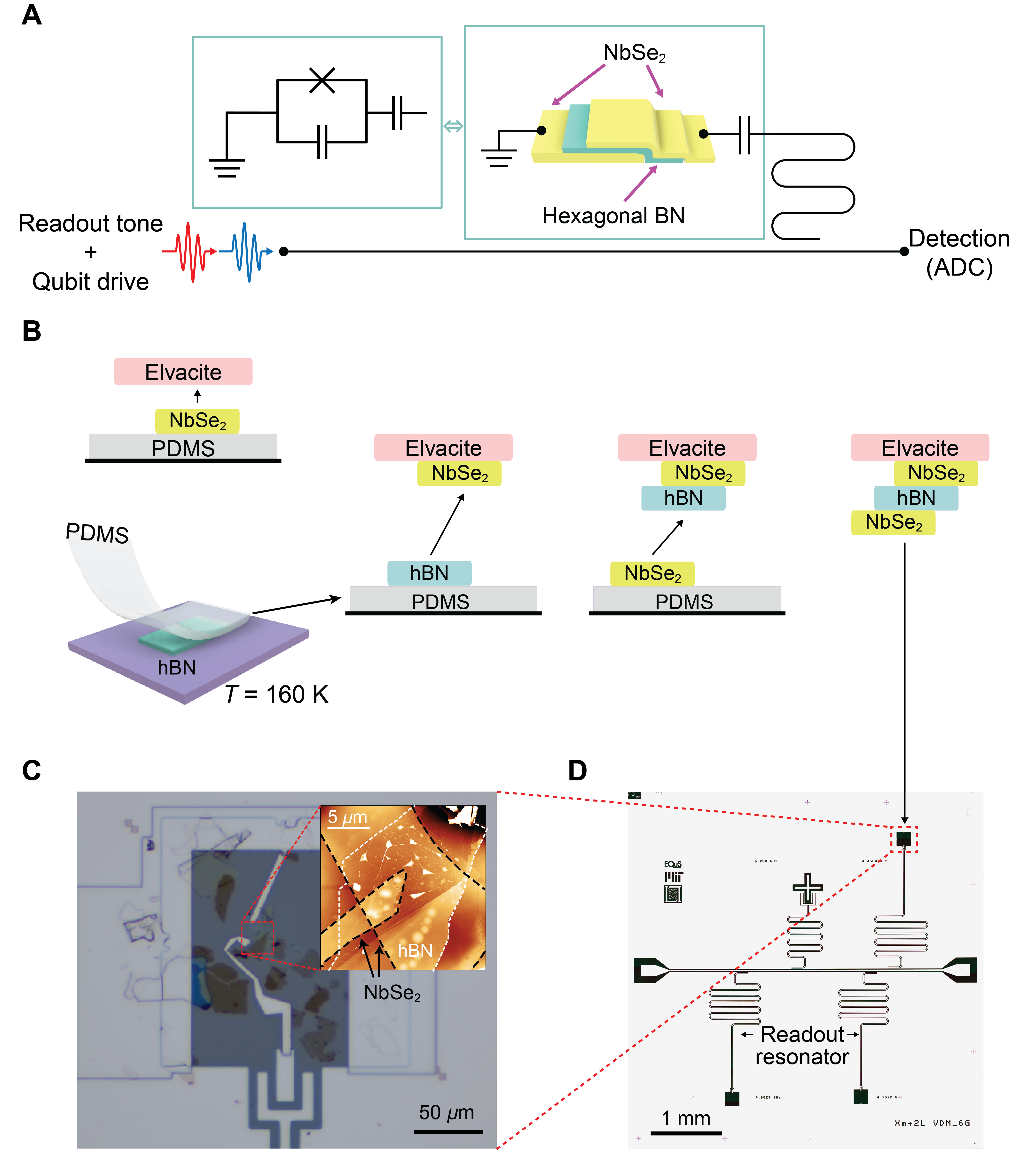}
    \caption{\textbf{Fabrication and circuit layout of an all-vdW, merged-element transmon qubit.}
    (\textbf{A}) Schematic of the circuit-QED architecture integrating the all-vdW, merged-element transmon qubit. 
    The resonator is inductively coupled to the feedline, while the NbSe$_2$--hBN--NbSe$_2$ heterostructure is capacitively coupled to the resonator. 
    (\textbf{B}) Schematic illustration of the polymer-based dry-transfer process. 
    A thin hBN flake is first picked up at \SI{160}{K} using a PDMS stamp and then flipped so that the hBN surface faces upward. 
    NbSe$_2$ and hBN flakes are sequentially stacked using a polymer (Elvacite) pick-up method to form the NbSe$_2$--hBN--NbSe$_2$ junction. 
    (\textbf{C}) Optical micrograph of an all-vdW, merged-element transmon qubit. Inset is an AFM image of the junction before the fabrication process. White and black dashed lines indicate the hBN and NbSe$_2$ flakes, respectively. 
    (\textbf{D}) Optical micrograph of a circuit-QED chip comprising three vdW qubits and one standard Al transmon~\cite{Schuster2007PhotonNumber,Schreier2008Suppressing,Xmon2013}.
    }
    \label{Figure1}
\end{figure}

A central challenge in modern quantum computing is integrating low-error-rate physical qubits into compact, scalable architectures that can support the large qubit counts required for quantum error correction~\cite{GoogleQuantumAI2025SurfaceCodeThreshold,Campbell2017Roads}.
Superconducting qubits---a leading platform to date---must be integrated into sufficiently large arrays to realize quantum error correction and thereby achieve fault-tolerant operation~\cite{Campbell2017Roads}. In these devices, dielectric loss is widely recognized as a primary limitation to coherence. Efforts to mitigate this loss have focused on reducing the participation of lossy surface and interfacial regions, where two-level systems (TLS) can form and couple dissipatively to qubits~\cite{Bland2025Millisecond, Place2021NewMaterialPlatform, Ofek2016ExtendingLifetime}. These strategies have shaped state-of-the-art superconducting multi-qubit processors, which typically rely on planar geometries and distributed circuit elements, resulting in single-qubit footprints on the order of $(300~\si{\micro\meter})^2$. Van der Waals (vdW) materials provide a means to realize lumped-element devices with significantly reduced form factor, and the challenge is to do so while maintaining high coherence and ultimately manufacturability~\cite{wang2019coherent,wang2022hexagonal, Jesse2026WSe2,Antony2021hBN,Balgley2025Coherent,DElia2025Coherent}.

VdW materials and heterostructures have attracted substantial interest from both the two-dimensional materials and superconducting-qubit communities~\cite{wang2019coherent, wang2022hexagonal, Antony2021hBN,Jesse2026WSe2,Balgley2025Coherent,DElia2025Coherent,Dean2010,Geim2013}. Ideally, their two-dimensional crystalline structure, atomically flat surfaces, and well-defined interfaces---combined with deterministic heterostructure assembly---also provide a means to assess the relative importance of crystalline materials for improving superconducting qubit performance~\cite{wang2019coherent, wang2022hexagonal, Antony2021hBN,Jesse2026WSe2,Balgley2025Coherent,DElia2025Coherent,Dean2010,Geim2013,Schmidt2018,Kroll2018}. 


Hexagonal boron nitride (hBN), a key component of vdW heterostructures, exhibits low microwave dielectric loss~\cite{wang2022hexagonal,Antony2021hBN}. Measurements at low temperatures and in the single-photon regime indicate a loss tangent $\tan\delta_{\mathrm{hBN}} < 10^{-5}$~\cite{wang2022hexagonal,Antony2021hBN}, more than two orders of magnitude below the typical loss tangent of deposited amorphous oxides~\cite{Deng2014AlOxLoss}. Parallel-plate capacitors formed from relatively thick ($>$\SI{10}{nm}) hBN layers and NbSe$_2$ flakes can reduce the footprint of a transmon shunt capacitor by more than two orders of magnitude by achieving participation ratios exceeding 90$\%$ within the hBN. These properties make hBN-based heterostructures potentially attractive for low-loss, compact superconducting quantum circuits. Experimentally, however, demonstrating an all-vdW superconducting qubit with coherence on par with conventional Al--AlO$_\mathrm{x}$--Al devices has remained an outstanding challenge.

Here, we realize an all-vdW qubit based on a NbSe$_2$--hBN--NbSe$_2$ heterostructure serving as the qubit within a circuit-QED architecture. A 2--3-layer hBN barrier is used to realize a superconducting transmon qubit with a frequency in the few-gigahertz range. In this merged-element geometry, a single overlap area serves simultaneously as the Josephson tunnel junction and shunt capacitor, providing the required Josephson coupling and capacitance within a compact footprint~\cite{zhao2020merged,mamin2021merged,Goswami2022FinMET,Jesse2026WSe2,Balgley2025Coherent,DElia2025Coherent}.
\section*{Device Fabrication and Circuits}


Figure~\ref{Figure1}\textbf{A} shows the circuit-QED architecture. The vdW qubit is capacitively coupled to a readout resonator, which is inductively coupled to the feedline. Qubit readout and control are performed by applying microwave drive and readout tones to the feedline and monitoring the transmitted signal. The detailed circuitry is described in Supplementary Fig. S1.

We assemble the NbSe$_2$--hBN--NbSe$_2$ vdW heterostructures using a polymer-based dry-transfer technique~\cite{Wang2013,wang2022hexagonal} (Fig.~\ref{Figure1}\textbf{B}). A three-layer hBN flake is sandwiched between two NbSe$_2$ flakes. To overcome weak adhesion between NbSe$_2$ and hBN, we use a cryogenic pick-up method~\cite{zhao2023Science}. A thin hBN flake exfoliated onto a SiO$_2$/Si wafer is first picked up at \SI{160}{K} using a polydimethylsiloxane (PDMS) stamp. In parallel, NbSe$_2$ flakes exfoliated onto PDMS are retrieved using an Elvacite polymer stamp~\cite{Elvacite2018}. The Elvacite/NbSe$_2$ stack is then brought into contact with the hBN on PDMS, allowing the hBN and NbSe$_2$ flakes to be picked up sequentially and forming the NbSe$_2$--hBN--NbSe$_2$ stack. Finally, the completed heterostructures are aligned and released onto the Al-based circuit-QED silicon chips shown in Fig.~\ref{Figure1}\textbf{D}. All exfoliation and transfer steps are performed in an argon atmosphere to preserve the NbSe$_2$ and hBN interfaces.

Among the fabricated devices, we focus on the vdW qubit shown in Fig.~\ref{Figure1}\textbf{C}, which corresponds to the heterostructure highlighted in Fig.~\ref{Figure1}\textbf{A}. All measurements reported in the main text were performed on this device.
We also fabricated a frequency-tunable device using a SQUID configuration, shown in Supplementary Figs. S2 and S3; parameters for the measured devices are summarized in Supplementary Table S2.

\section*{Qubit spectroscopy and coherent control}
\begin{figure}[htbp]
    \centering
    \includegraphics[width=1\linewidth]{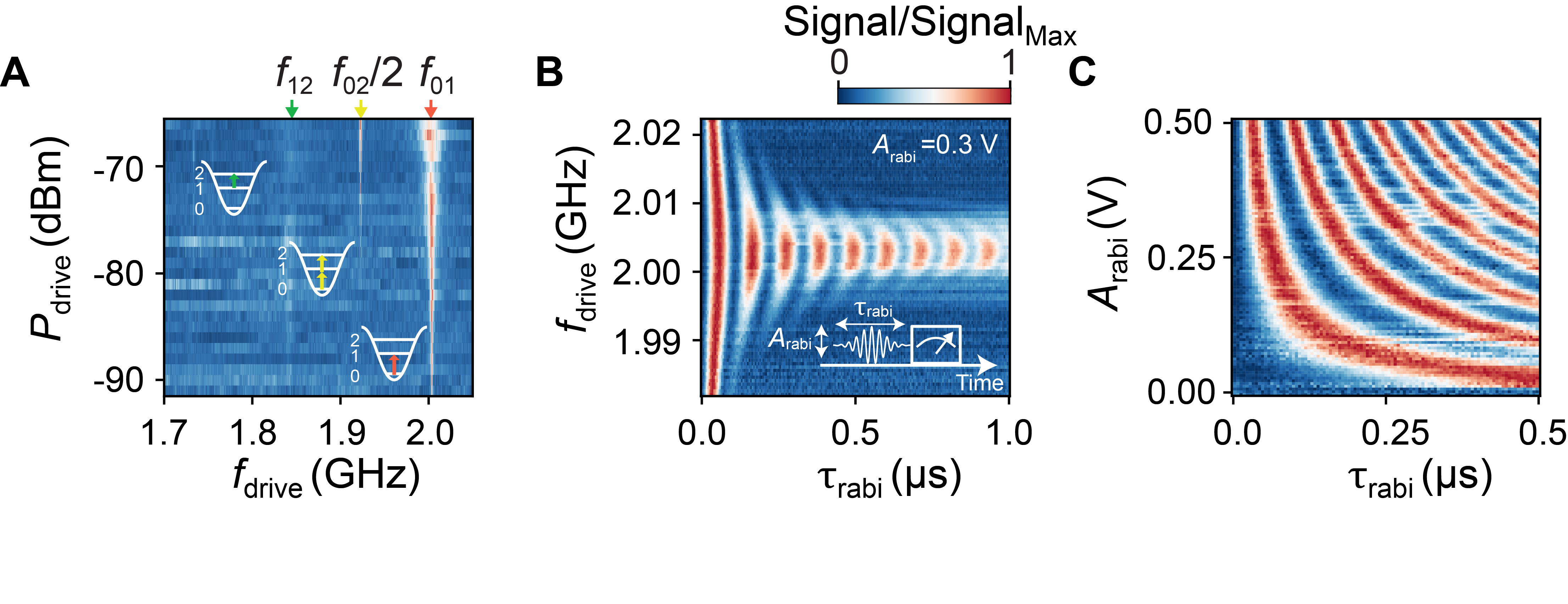}
    \caption{\textbf{Qubit spectroscopy and coherent control of the all-vdW merged-element transmon qubit.} 
    (\textbf{A}) Two-tone spectroscopy as a function of the drive power and drive frequency to identify the qubit transition frequencies. 
    The fundamental transition $f_{01}$ appears at \SI{2.00}{\giga\hertz} (red arrow), the two-photon transition $f_{02}/2$ at \SI{1.92}{\giga\hertz} (yellow arrow), and the first-excited-state transition $f_{12}$ at \SI{1.84}{\giga\hertz} (green arrow). 
    As the drive power increases, the spectral peaks broaden and their amplitudes increase, indicating stronger excitation of higher-order transitions. The
intermediate-frequency (IF) signal amplitude is fixed at
\SI{1}{\volt}, and the drive power ($P_{\mathrm{drive}}$) at the device is estimated from the total attenuation
of the qubit-drive line. 
    (\textbf{B}) Rabi oscillations measured at an IF drive amplitude of \SI{0.3}{V}. The microwave signal is upconverted using a local
oscillator (LO) power of $15~\mathrm{dBm}$. Clear oscillations are observed as a function of drive-pulse duration. 
    (\textbf{C}) Rabi oscillations measured at $f_{\mathrm{drive}}=f_{01}$  as a function of both drive amplitude and pulse duration, illustrating coherent evolution of the qubit state. The LO power is fixed at 15 dBm. The quoted drive amplitudes and power correspond to room-temperature settings.} 
    \label{figure2}
\end{figure}

To characterize the fabricated vdW qubit, we first perform standard two-tone spectroscopy by applying a drive tone and then monitoring the transmitted readout signal. The readout tone is fixed at the resonator frequency, $f_{\mathrm{res}}$, while the drive frequency is swept.
Figure~\ref{figure2}\textbf{A} shows the change in the readout signal as a function of the drive frequency $f_{\mathrm{drive}}$ and drive power $P_{\mathrm{drive}}$. At low drive powers, the spectrum exhibits a sharp resonance at $f_{\mathrm{drive}} = \SI{2.00}{\giga\hertz}$, which we identify as the qubit transition frequency $f_{01}$. As the drive power increases, the peak at $f_{01}$ broadens and a second peak appears at $f_{02}/2 = \SI{1.92}{\giga\hertz}$, corresponding to a two-photon transition to the second excited state of the qubit.
We fit the observed transitions within the Cooper-pair-box model~\cite{Krantz2019QuantumEngineersGuide,Kyle2019},
\begin{equation}
H = 4E_{\mathrm{C}}\left(\hat{n}-n_{\mathrm{g}}-\frac{P-1}{4}\right)^2 - E_{\mathrm{J}} \cos \hat{\phi},
\label{QubitHamiltonian}
\end{equation}
where $\hat{n}$ is the dimensionless Cooper-pair number operator, $n_{\mathrm{g}}$ is the dimensionless offset charge normalized to $2e$, $P=\pm1$ denotes the charge-parity state, and $\hat{\phi}$ is the phase operator of the Josephson junction. Fitting the measured transition frequencies yields a Josephson energy $E_{\mathrm{J}}/h = \SI{4.29}{\giga\hertz}$ and a charging energy $E_{\mathrm{C}}/h = \SI{0.13}{\giga\hertz}$, giving $E_{\mathrm{J}}/E_{\mathrm{C}} \approx 33$. For these parameters, the calculated charge dispersion of the qubit transition, $\delta f_{01} = |f_{01}(n_{\mathrm{g}} = 0) - f_{01}(n_{\mathrm{g}} = 0.5)|$, is approximately \SI{160}{\kilo\hertz} (see Table~\ref{Qubit_params}). The detailed Hamiltonian and fitting process are described in Supplementary Sec.~6.

\begin{table}[t]
\centering
\caption{Extracted circuit and dispersive-readout parameters}
\label{Qubit_params}
\begin{tabular}{l c}
\hline\hline
Parameter & Value \\
\hline
Resonator frequency, $f_{\mathrm{res}}$ & \SI{6.27}{\giga\hertz}\\
Qubit transition frequency, $f_{01}$ & \SI{2.00}{\giga\hertz}\\
Josephson energy, $E_{\mathrm{J}}/h$ & \SI{4.29}{\giga\hertz} \\
Charging energy, $E_{\mathrm{C}}/h$ & \SI{0.13}{\giga\hertz} \\
Charge dispersion, $\delta f_{01}$ & \SI{160}{\kilo\hertz} \\
Resonator linewidth, $\kappa/2\pi$ & \SI{454}{\kilo\hertz} \\
Lamb shift (referenced to ground), $\chi_{0}/2\pi$ & \SI{155}{\kilo\hertz} \\
Dispersive shift between $|0\rangle$ and $|1\rangle$, $\chi_{01}/2\pi$ & \SI{34}{\kilo\hertz} \\
\hline\hline
\end{tabular}
\end{table}


We next use Rabi oscillations to calibrate the  amplitude and duration of $\pi$ and $\pi/2$ pulses and establish coherent control of the vdW qubit. The RF drive tone has a cosine envelope with amplitude $A_{\mathrm{rabi}}$ and duration $\tau_{\mathrm{rabi}}$. Figure~\ref{figure2}\textbf{B} shows the measured readout signal as a function of drive frequency $f_{\mathrm{drive}}$ and pulse duration $\tau_{\mathrm{rabi}}$ at a fixed drive amplitude $A_{\mathrm{rabi}}=\SI{0.3}{\volt}$.  
Applying a pulse near the qubit transition frequency induces coherent Rabi oscillations between the ground $|0\rangle$ and excited $|1\rangle$ states. At resonance, we obtain the largest oscillation amplitude and the minimum Rabi frequency $\Omega_{\mathrm{rabi}}/2\pi=\SI{1.46}{\mega\hertz}$. As the drive is detuned from $f_{01}$, the oscillation amplitude decreases and the Rabi frequency increases (fringes become more closely spaced), producing the characteristic chevron pattern. 
Next, we repeat the Rabi measurement with the drive tone frequency fixed at the qubit frequency, $f_{\mathrm{drive}}=f_{01}$, while varying  $\tau_{\mathrm{rabi}}$ and $A_{\mathrm{rabi}}$ [Fig.~\ref{figure2}\textbf{C}].
As expected, the Rabi oscillation frequency increases linearly with the pulse amplitude.

\begin{figure}[htbp]
    \centering
    \includegraphics[width=1.0 \linewidth]{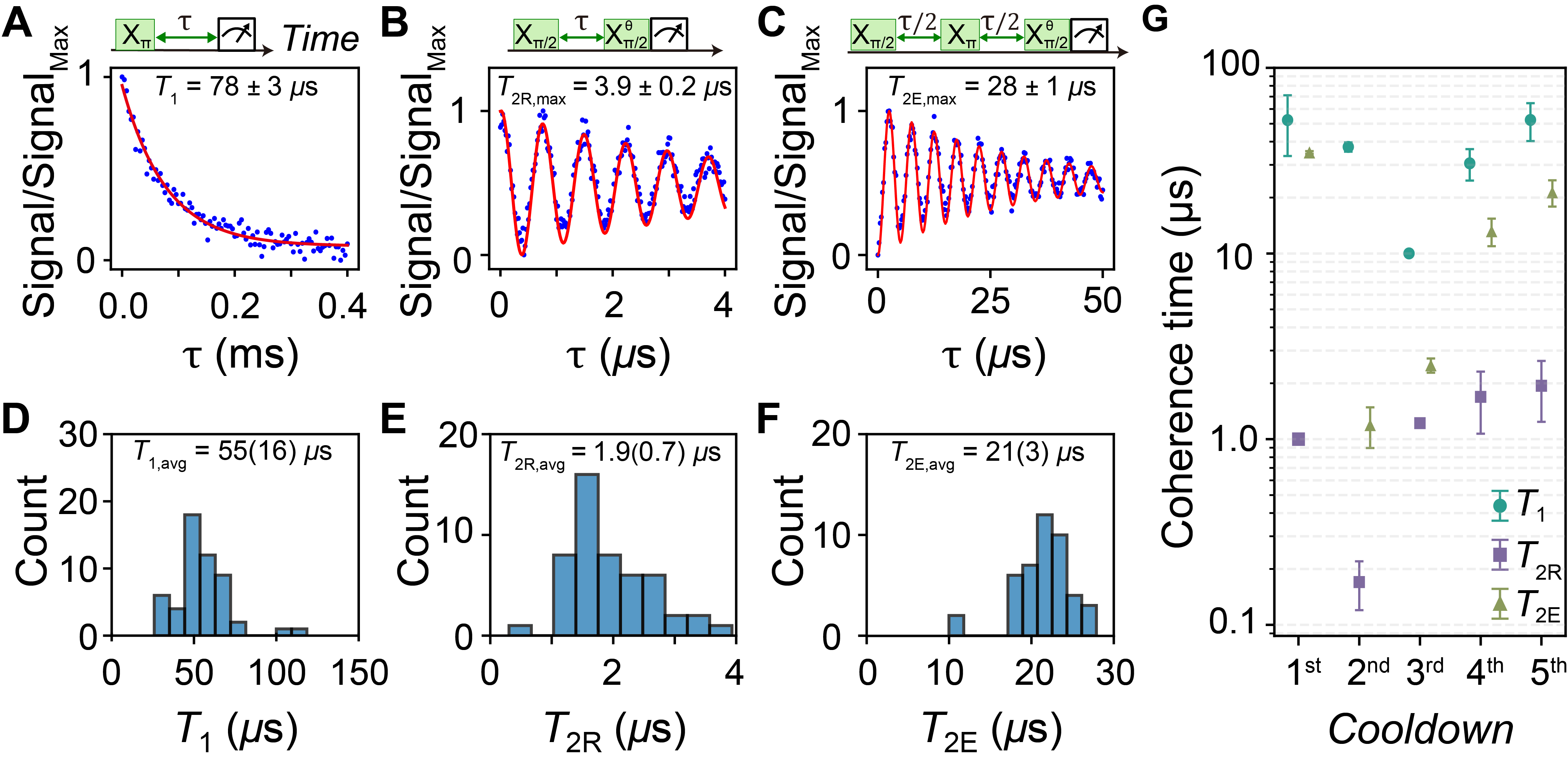}
        \caption{\textbf{Qubit coherence properties.}
        (\textbf{A}) The relaxation time \(T_1\) is extracted from an exponential fit (red line) to the normalized signal as a function of measurement delay, yielding \(T_1=78\pm3~\mu\text{s}\). The trace shown is representative of the distribution in panel D, and values following $\pm$ represent the fit uncertainty.
        (\textbf{B}) Ramsey dephasing experiment, consisting of two $\pi/2$ control pulses separated by a delay. The phase of the second pulse is advanced by $\tau$ - proportional shift relative to the first, producing decaying oscillations with a characteristic dephasing time $T_{2\mathrm{R,max}} = 3.9 \pm 0.2~\si{\micro\second}$. The trace shown is the maximum value of the distribution in Panel E.
        (\textbf{C}) Hahn-echo dephasing experiment, $\pi/2-\pi-\pi/2$ pulse sequence with a $\tau$ - proportional shift relative to the final pulse yields an echo-limited dephasing time $T_{2\mathrm{E,max}} = 28 \pm 1~\si{\micro\second}$. The trace shown is the maximum value of the distribution in Panel F.
        (\textbf{D}) Histogram of relaxation times measured continuously over 10 hours. The distribution has an average of  $T_{1,\mathrm{avg}} = 55~\si{\micro\second}$, with a standard deviation of $16~\si{\micro\second}$. 
        (\textbf{E--F}) Histograms of interleaved Ramsey and Hahn-echo dephasing times measured over 10 hours. The dephasing time distributions have average values of Ramsey and Hahn-echo dephasing times of $T_{2\mathrm{R},\mathrm{avg}} = 1.9 ~\si{\micro\second}$, and $T_{2\mathrm{E},\mathrm{avg}} = 21~\si{\micro\second}$, with standard deviations of $0.7~\si{\micro\second}$ and $3~\si{\micro\second}$, respectively.
        (\textbf{G}) Relaxation and coherence times ($T_{1}$, $T_{2\mathrm{R}}$, and $T_{2\mathrm{E}}$) as a function of cooldown number.
}

    \label{figure3}
\end{figure}

Having established temporal control of the qubit state, we next characterize temporal relaxation and dephasing, which determine the qubit coherence limits and probe noise sources in the vdW qubit and its environment. We measure the energy-relaxation time $T_1$ using an inversion-recovery sequence: a calibrated $\pi$ pulse prepares the qubit in $|1\rangle$, followed by a variable delay time $\tau$ and projective readout. The excited-state population decays exponentially with a characteristic time $T_1 = 78 \pm 3~\si{\micro\second}$ [Fig.~\ref{figure3}\textbf{A}]. Across 50 consecutive measurements acquired over a 10-hour period, the distribution of fitted $T_1$ values has an average of $T_{1,\mathrm{avg}} = 55~\si{\micro\second}$, with a standard deviation of $16~\si{\micro\second}$ [Fig.~\ref{figure3}\textbf{D}].

We probe dephasing using Ramsey and Hahn-echo sequences. In Ramsey interferometry, we apply a resonant $\pi/2$ pulse, allow free evolution for a variable delay time $\tau$, and then apply a second $\pi/2$ pulse with a digitally rotated phase to introduce a small effective detuning. The resulting interference fringes decay with a time constant $T_{2\mathrm{R}} = 3.9 \pm 0.2~\si{\micro\second}$. 
Using a Hahn-echo sequence, in which an additional $\pi$ pulse is inserted midway through the free-evolution period, we measure a coherence time $T_{2\mathrm{E}} = 28 \pm 1~\si{\micro\second}$. 
Interleaved Ramsey and Hahn-echo measurements across 50 repetitions over the same 10-hour period yield distributions of fitted coherence times with average values $T_{2\mathrm{R},\mathrm{avg}} = 1.9~\si{\micro\second}$ and $T_{2\mathrm{E},\mathrm{avg}} = 21~\si{\micro\second}$, and standard deviations of $0.7~\si{\micro\second}$ and $3~\si{\micro\second}$, respectively.

Using the Bloch-Redfield relation
\[
\frac{1}{T_2}=\frac{1}{2T_1}+\frac{1}{T_\phi},
\]
we extract pure dephasing times from the averaged coherence metrics. We obtain $T_{\phi\mathrm{R},\mathrm{avg}} = 1.9~\si{\micro\second}$ and $T_{\phi\mathrm{E},\mathrm{avg}} = 26 ~\si{\micro\second}$, and propagated standard deviations of $0.7~\si{\micro\second}$ and $5~\si{\micro\second}$, respectively. Although the Bloch-Redfield equation is not strictly valid for 1/f noise~\cite{Ithier2005Decoherence,Bylander2011Noise,Yoshihara2006Decoherence}, the pronounced difference between Ramsey and echo indicates that dephasing is dominated by low frequency noise that is efficiently refocused by the Hahn-echo sequence. The fact that $T_{2\mathrm{E}}$ remains well below the $2T_1\approx 100 ~\si{\micro\second}$ limit suggests that additional non-refocusable dephasing channels at higher frequency also contribute to the overall decoherence.
Figure~\ref{figure3}\textbf{G} shows the extracted $T_1$, $T_{2\mathrm{R}}$, and $T_{2\mathrm{E}}$ across multiple cooldowns. The relaxation and dephasing times fluctuate from cooldown to cooldown, as is also observed with conventionally fabricated superconducting qubits~\cite{Burnett2019Decoherence,Klimov2018Fluctuations,Schlor2019Correlating}, with the longest $T_1$ observed in the fifth cooldown. Such cooldown-to-cooldown variability may arise from changes in the TLS environment coupled to the qubit~\cite{Muller2015Interacting}.

\section*{Even--odd charge parity and coherence}

\begin{figure}[htbp]
    \centering
    \includegraphics[width=0.8\linewidth]{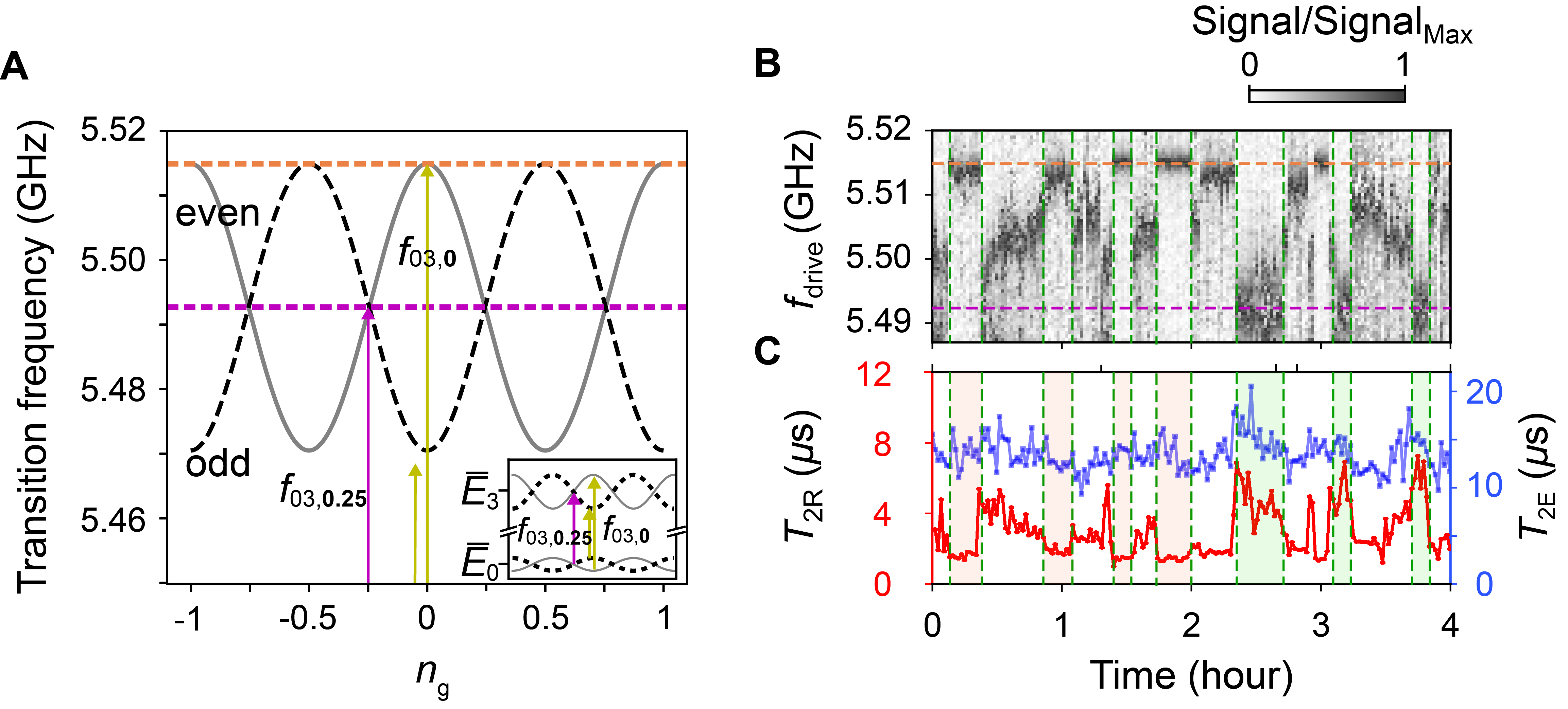}
    \caption{\textbf{Offset-charge dependence of qubit transition frequency in an all-vdW transmon qubit.} 
    (\textbf{A}) Calculated transition frequency between the ground and third excited states as a function of offset charge $n_{\mathrm{g}}$ (in units of $2e$), based on the extracted $E_{\mathrm{J}}$ and $E_{\mathrm{C}}$. 
Charge parity (even/odd) shifts the transition band, producing parity-sensitive (orange dashed line) and parity-insensitive (purple dashed line) operating points. 
Inset: energy levels of the ground and third excited states, with $E_0$ and $E_3$ indicating time-averaged energies. 
    (\textbf{B}) Repeated two-tone spectroscopy of the ground-to-third-state transition. 
The orange and purple dashed lines correspond to the operating points marked in (A); the purple dashed line highlights transitions through parity-insensitive regions.
    (\textbf{C}) Repeated Ramsey dephasing and Hahn-echo measurements interleaved with the spectroscopy in (B), revealing that Ramsey dephasing is quasiparticle-limited. The measurements were performed during the fourth cooldown.}
    \label{Figure4}
\end{figure}

The disparity between the Ramsey and Hahn-echo coherence times, $T_{2\mathrm{R}} \ll T_{2\mathrm{E}}$, indicates that dephasing is dominated by low-frequency fluctuations of the qubit transition frequency. A natural microscopic source of such fluctuations is charge dynamics arising from the finite charge dispersion of the device. 
Because the qubit has a relatively low Josephson-to-charging energy ratio ($E_{\mathrm{J}}/E_{\mathrm{C}} \approx 33$), it is not fully in the charge-insensitive transmon limit and instead retains measurable offset-charge sensitivity. In this regime, fluctuations of the offset charge $n_{\mathrm{g}}$ and quasiparticle-induced charge-parity switching can both shift the qubit transition frequency~\cite{Koch2007Transmon}, producing the low-frequency dephasing noise inferred from the Ramsey and Hahn-echo measurements. The relative importance of these mechanisms depends on the offset-charge operating point (Supplementary Fig. S5). Near $n_{\mathrm{g}} = 0$, the qubit frequency is primarily sensitive to even--odd charge parity, where stochastic quasiparticle tunneling induces telegraph-like frequency jumps between the two parity branches. Near $n_{\mathrm{g}} = 0.25$, by contrast, the even- and odd-parity branches become nearly degenerate while the transition frequency has a maximum slope with respect to $n_{\mathrm{g}}$, making the qubit especially sensitive to offset-charge fluctuations.

Motivated by this expected charge sensitivity, we perform interleaved measurements of the ground-to-third-excited-state transition frequency $f_{03}$, Ramsey coherence time $T_{2\mathrm{R}}$, and Hahn-echo coherence time $T_{2\mathrm{E}}$ to directly correlate dephasing with offset charge. The $f_{03}$ transition has substantially larger charge dispersion than the lower transitions and therefore serves as an \textit{in situ} monitor of offset-charge fluctuations and quasiparticle-induced parity switching during the time-domain measurements (Supplementary Fig. S5).

Figure~\ref{Figure4}\textbf{A} shows the calculated $f_{03}(n_{\mathrm{g}})$ dispersion based on the extracted $E_{\mathrm{J}}$ and $E_{\mathrm{C}}$, where the transition splits into two branches associated with different charge-parity manifolds (even/odd). For our device ($E_{\mathrm{J}}/E_{\mathrm{C}}\approx 33$), the integer-charge point $n_{\mathrm{g}}=0$ corresponds to $f_{03}=\SI{5.515}{\giga\hertz}$ (orange dashed line) for the even-parity state, with a maximum parity splitting (charge dispersion) of $\delta f_{03}=\SI{44}{\mega\hertz}$.
In contrast, near $n_{\mathrm{g}}\approx 0.25$, the two parity branches approach a crossing point at $f_{03}=\SI{5.492}{\giga\hertz}$ (purple dashed line), where the transition becomes insensitive to parity switching but remains sensitive to fluctuations in $n_{\mathrm{g}}$~\cite{Koch2007Transmon,Kyle2018,Kyle2019,Riste2013MillisecondChargeParity}.
In Fig.~\ref{Figure4}\textbf{B} (top panel), repeated spectroscopy measurements show that the $|0\rangle\rightarrow|3\rangle$ transition frequency fluctuates over time due to slow offset-charge drift (for example, in the substrate). Note that, to increase the measurement repetition rate, we restrict the frequency sweep only to the upper branch of the spectrum (between the purple and orange dashed lines in Fig.~\ref{Figure4}\textbf{A}). The observed frequency fluctuations are in good agreement with the calculated charge dispersion.

The interleaved data reveal the expected strong correlation between $T_{2\mathrm{R}}$ and $n_{\mathrm{g}}$: near $n_{\mathrm{g}} \approx 0.25$ the Ramsey time $T_{\mathrm{2R}}$ increases to $\sim 6~\si{\micro\second}$, whereas near $n_{\mathrm{g}}\approx 0$ or $0.5$ it remains short, $T_{2\mathrm{R}}\sim 1$--$2~\si{\micro\second}$ (Supplementary Fig. S6). 
Notably, the suppression of $T_{2\mathrm{R}}$ at $n_{\mathrm{g}}=0$ coincides with the point of maximum parity splitting, indicating that Ramsey coherence is limited by quasiparticle tunneling.
Conversely, near $n_{\mathrm{g}}\approx 0.25$, where the parity branches are nearly degenerate, parity switching induces only a weak instantaneous frequency shift and Ramsey coherence is correspondingly enhanced. The observed $T_{2\mathrm{R}}\approx 6~\si{\micro\second}$ remains about a factor of two shorter than $T_{2\mathrm{E}}$, consistent with residual low-frequency dephasing from other sources~\cite{Krantz2019QuantumEngineersGuide,Ithier2005Decoherence}. 
Although the strong correlation between $T_{2\mathrm{R}}$ and $n_{\mathrm{g}}$ explains the short Ramsey coherence in terms of quasiparticle tunneling, the nonideal $T_{2\mathrm{E}}$ is not explained by this mechanism. $T_{2\mathrm{E}}$ shows no strong dependence on offset charge, suggesting that neither quasiparticle tunneling nor offset-charge fluctuations are the dominant limitations of $T_{2\mathrm{E}}$ (Supplementary Fig. S6). The measured $T_{2\mathrm{E}}$ is still well below the relaxation-limited value, indicating an additional dephasing channel (or channels). Possible mechanisms include photon shot noise in the resonator, excess thermal photons in the qubit control lines or environment, critical current fluctuations in the junction, and fluctuations associated with microscopic two-level systems or defects. We estimate the photon-shot-noise contribution using the measured qubit and resonator parameters. Explaining the observed echo dephasing would require an effective resonator temperature of approximately 636 mK, far above the mixing chamber temperature or the effective resonator temperatures observed in other experiments from our group~\cite{Jin2015Thermal}, making this mechanism unlikely to dominate $T_{2\mathrm{E}}$ (Supplementary Fig. S7). Similarly, the control lines and qubit environment are nominally identical to other superconducting qubit experiments performed by the group and therefore thermal photons are not a likely candidate at this level of dephasing. Critical-current fluctuations induced by microscopic defects in the tunnel barrier or nearby interfaces remain possible sources of residual echo dephasing and may be reduced through future improvements in junction quality and interface control (Supplementary Fig. S8).

\section*{Conclusion}
In this work, we realized an all-vdW merged-element transmon qubit based on a NbSe$_2$--hBN--NbSe$_2$ heterostructure serving as the qubit element within a circuit-QED architecture. By using a thin 2--3-layer hBN barrier that simultaneously serves as the Josephson tunnel barrier and the shunt-capacitor dielectric, the electric field is strongly confined to a small volume of relatively high-quality dielectric, enabling a compact qubit footprint. 

We demonstrate coherent control through Rabi measurements and quantify the coherence properties via relaxation and dephasing experiments. The representative relaxation time shows $T_1 = 78 \pm 3~\si{\micro\second}$. Across repeated measurements, the distribution of the fitted relaxation times has an average of $T_{1,\mathrm{avg}} = 55~\si{\micro\second}$ with a standard deviation of $16~\si{\micro\second}$. Dephasing measurements reveal a large separation between Ramsey and Hahn-echo coherence ($T_{2\mathrm{R},\mathrm{avg}}\approx 1.9~\si{\micro\second}$ versus $T_{2\mathrm{E},\mathrm{avg}}\approx 21~\si{\micro\second}$), indicating that low-frequency noise plays a dominant role and can be refocused to a large degree, but not enough for $T_\mathrm{2}$ to reach the $2T_1$ limit. Given the relatively low $E_{\mathrm{J}}/E_{\mathrm{C}}\approx33$ ratio in this device, charge noise is a likely contributor to the observed dephasing. We confirm this with interleaved $f_{03}$ spectroscopy and Ramsey measurements and observe a strong dependence of $T_{2\mathrm{R}}$ on offset charge: $T_{2\mathrm{R}}$ improves to $\sim 6~\si{\micro\second}$ near $n_{\mathrm{g}}\approx 0.25$ but remains short ($\sim 1$--$2~\si{\micro\second}$) near $n_{\mathrm{g}}\approx 0$ and $0.5$. This correlation is consistent with quasiparticle-induced charge-parity switching, which produces telegraph-like frequency fluctuations that strongly suppress Ramsey coherence at parity-sensitive operating points, while dephasing is reduced near the parity-insensitive point.

These results establish vdW heterostructures as a viable platform for realizing superconducting qubits. The observed charge-noise-induced dephasing is not fundamental to vdW qubits, but instead arises from a relatively low $E_{\mathrm{J}}/E_{\mathrm{C}}$ ratio. Future improvements include device-level optimization to further suppress charge dispersion, for example by increasing the junction area to raise $E_{\mathrm{J}}/E_{\mathrm{C}}$ above 50, as well as improved interface cleanliness and materials processing. In particular, topographic features consistent with microscopic residues or transfer-related contamination are visible in the AFM image in Fig.~\ref{Figure1}\textbf{C}, indicating clear opportunities for fabrication improvements that may further reduce loss and noise. More broadly, once these issues associated with manufacturing and cleanliness are resolved, the deterministic stacking and materials modularity of vdW heterostructures provide a promising route for exploring new superconducting circuit elements and scalable circuit-QED implementations based on atomically engineered materials.




\bibliographystyle{sciencemag}
\bibliography{vdW_qubit}


\section*{Acknowledgments}
We thank Renée DePencier Piñero, Kevin Grossklaus, Xiaomeng Cui, Shuwen Sun, Xueqiao Wang, Ruihao Liu, Pablo Mercader-Pérez,  Ilan T. Rosen, Shantanu Jha, and the device packaging team at MIT Lincoln Laboratory for their assistance in measurement, fabrication, and packaging. This work was carried out in part using the MIT.nano's facilities.

\paragraph*{Funding:}
This research was funded in part by the US Army Research Office grant No.~W911NF-22-1-0023, by the  National Science Foundation under grants No. PHY-2412810 and No.~OMA-1936263, by the Air Force Office of Scientific Research grant No.~FA2386-21-1-4058, and under Air Force Contract No.~FA8702-15-D-0001.
S.P. was supported by the Education and Training Program of the Quantum Information Research Support Center (No. RS-2026-25501060), the National Research Foundation of Korea (NRF) (No. RS-2021-NR060139), and the Soseon Foundation.
S.P. and G.-H.L. were supported by the NRF (No. RS-2025-02317602, No. RS-2024-00393599) and by the ITRC (Information Technology Research Center) support program (No. IITP-2026-RS-2022-00164799).
S.Z. acknowledges support from the Faculty for the Future Fellowship from the Schlumberger Foundation.
J.A. and J.K. acknowledge support from Korea Foundation for Advanced Studies (KFAS). C.-C.T. acknowledges support from the Think Global fellowship.
D. R-L. acknowledges support from the Rafael del Pino Foundation.
K.W. and T.T. acknowledge support from the JSPS KAKENHI (Grant Numbers 21H05233 and 23H02052), the CREST (JPMJCR24A5), JST, and the World Premier International Research Center Initiative (WPI), MEXT, Japan.
P.J.-H. acknowledges support by the Air Force Office of Scientific Research (AFOSR) grant FA9550-21-1-0319, the Office of Naval Research (ONR) grant N000142412440, the MIT Microsystems Technology Laboratories, Samsung Semiconductor Research Fund, the Gordon and Betty Moore Foundation’s EPiQS Initiative through Grant No. GBMF9463, the Fundacion Ramon Areces, and the CIFAR Quantum Materials program.
Any opinions, findings, conclusions, or recommendations expressed in this material are those of the author(s) and do not necessarily reflect the views of the U.S. Air Force or the U.S. Government.

\paragraph*{Author contributions:}

J.~${\hat{I}}$-J.~W. conceived and designed the experiment. S.P., J.~${\hat{I}}$-J.~W., and S.Z. performed the microwave simulations. S.P., J.~${\hat{I}}$-J.~W., S.Z., D.R.-L., H.-Y.T., C.-C.T., and A.G. contributed to the device fabrication. S.P., S.Z., J.K., and J.A. participated in the measurements. S.P., J.K., and J.A. analyzed the data. K.W. and T.T. grew the hBN crystal. S.P., S.Z., J.K., J.A., R.A., W.P.B., and G.D.C. contributed to the measurement setup. S.P., J.~${\hat{I}}$-J.~W., and W.D.O. led the paper writing, with contributions from all authors. T.P.O., G.-H.L., K.S., M.H., J.A.G., P.K., P.J.-H., J.~${\hat{I}}$-J.~W., and W.D.O. supervised the project.

\paragraph*{Competing interests:}
The authors declare no competing interests.
\paragraph*{Data and materials availability:}
The data that support the findings of this study are available from the corresponding authors upon reasonable request and with the cognizance of our US government sponsors who funded the work.

\end{document}